\documentclass[twocolumn,showpacs,amsmath,amssymb]{revtex4}

\usepackage{graphicx} % Include figure files
\usepackage{dcolumn}  % Align table columns on decimal point
\usepackage{bm}       % bold math
\newcommand{\lb}{{<}}
\newcommand{\rb}{{>}}

\begin{document}

\title{Kinetics of the Melting Transition in DNA}

\author{A. Santos}
\author{W. Klein}

\affiliation{Department of Physics, Boston University, Boston, MA 02215}

\date{\today}

\begin{abstract}
We investigate the kinetics of the DNA melting transition using modified versions of the Peyrard-Dauxois-Bishop and Poland-Scheraga models that include long and short range interactions.  
Using Brownian dynamics and Monte Carlo simulations, we observe metastable states prior to nucleation and demonstrate that the profile and growth modes of the critical droplet 
can have both classical and spinodal characteristics depending on the interaction range and the temperature quench depth.  
\end{abstract}


\maketitle

The melting of DNA from double to single-stranded form has been studied extensively for over forty years~\cite{Wartell, Poland}.  This transition continues to generate interest because it provides 
insight into the biological mechanisms of transcription and replication.   
Although much theoretical work has centered on the role of large nonlinear excitations as a precursor to melting~\cite{Englander,Dauxois,Dauxois2}, surprisingly little focus has been given to the kinetics 
of this peculiar one-dimensional phase transition. 

Nucleation, the process by which a system decays from a metastable state, plays an important role in many systems undergoing a phase transition~\cite{Zettlemoyer}.  
During nucleation, a critical droplet overcomes a free energy barrier and grows, causing the system to decay into the stable state~\cite{Langer}.  
DNA melting is believed to undergo such a process because its sharp melting curve indicates a first-order phase transition~\cite{Wartell} with 
the possibility of metastable states and because possible nucleation bubbles have been observed via electron microscopy~\cite{Pavlov}.  

Although the effect of including helicity~\cite{Barbi}, nonlinearities~\cite{Dauxois,Dauxois2}, sequence~\cite{Campa, Cule},
and defects~\cite{Singh} in models of DNA has been studied extensively, little has been done to study how long-range interactions affect these models.  
It is well known that systems with long-range interactions undergoing phase transitions can be quenched into metastable states near a 
pseudospinodal~\cite{Laradji, Herrmann, Klein}.  In the mean-field limit, this pseudospinodal 
becomes a well-defined spinodal~\cite{Herrmann, Klein, Gulbahce}, which is the limit of  metastability.  Systems undergoing spinodal nucleation are driven to the stable phase by diffuse fractal-like critical droplets whose 
amplitudes differ little from the metastable background~\cite{Unger, Unger2, Gould}.  In contrast, classical nucleation is initiated by compact droplets 
that resemble the stable phase.  
Hence, the inclusion of long-range interactions in models of DNA has implications for the nature of nucleation.  
It is also known that nearest-neighbor models of DNA typically underestimate the opening probability for small loops~\cite{Wartell, Blossey}, and experimental evidence suggests the interactions
may be long-range~\cite{Wartell2}.
In this Letter, we analyze how the range of interactions affects the kinetics and nature of the DNA melting transition in modified versions of the 
Peyrard-Dauxois-Bishop (PDB)~\cite{Dauxois,Dauxois2} and Poland-Scheraga (PS)~\cite{Poland} models.

In our modified version of the PDB model, the state of each base pair (bp) is specified by its 
separation $y_n$ and velocity $\dot{y}_n$.  
The Hamiltonian is given by $H_{\rm PDB} =  \sum_{i=1}^N m\dot{y}_i^2/2+V(y_i)$, 
where the first term is the kinetic energy of bps with mass $m=300$\,amu and the second term is a potential given by 
\begin{equation}
    \begin{split}
V(y_i) & =D(e^{-ay_i}-1)^2+ \sum_{j=i-R}^{i-1}W(y_i,y_j) \\
&-h(y_i-y_c)(T-T_c)\Theta(y_i-y_c).
\label{1}
    \end{split}
\end{equation}
The first term in Eq.~(\ref{1}) is an on-site Morse potential with dissociation energy $D=0.04$\,eV  and a parameter $a=4.45$\,\AA$^{-1}$ that goes as the inverse well width.  The Morse potential represents 
the attraction due to hydrogen bonds 
and the repulsion of negatively charged phosphate groups.  The term $W(y_n,\dot{y}_m)$ represents the stacking interaction and the elasticity of the phosphate backbone and is given by the 
anharmonic potential
\begin{equation}
W(y_n,\dot{y}_m)=\frac{1}{2}K\big(1+\rho e^{-\alpha (y_n+y_m)} \big)(y_n-y_m)^2.
\end{equation}
Here, the strength of the backbone is given by $K=0.06$\,eV/\AA$^{2}$, and the parameters $\alpha=0.35$\,\AA$^{-1}$ and $\rho=0.5$ represent the stacking interaction between bound bps.
The values of these parameters are similar to those in the literature~\cite{Dauxois,Dauxois2}.
In contrast to the PDB model, we let our interaction extend beyond nearest-neighbor to all bps within a range $R$, which is allowed to extend to sizes of order the 
persistence length of double-stranded DNA.  The persistence length, defined as the scale over which correlations in the tangential direction of the chain are lost, is of the order 100-200\,bps in 
double-stranded DNA.

The last term in Eq.~(\ref{1}) has been added to correct for entropic effects.  Simulations of the PDB model do not see separation of the strands in a reasonable 
time because the dissociated chain acts as a pseudorandom walker in one dimension, which without including higher spatial dimensions quickly returns to the origin even when running at temperatures well 
above the melting temperature.  Because this effect is not physical, the last term has been added to mimic the decrease in the free energy at large separations that would be felt by real DNA melting in 
solution.  The step function $\Theta(y_i-y_{c})$ turns on the potential when bps reach separations greater than a critical value $y_c=2.5$\,\AA, above which the bases are 
assumed to be in the stable open phase.
The interaction strength is given by the parameter $h=0.01$\,eV/\AA\  
and by the difference between the temperature $T$ and the assumed melting temperature $T_c=350$\,K.  

In the PS model, different statistical weights are given to bound and unbound segments.  A bound segment is energetically favored because of hydrogen bonding and stacking 
interactions.  Unbound segments are entropically favored because the single-stranded segments of a loop have a much shorter persistence length allowing them to sample a 
larger configuration of phase space.  Our modified PS model is described by 
the Hamiltonian
\begin{equation}
\label{3}
    \begin{split}
   	H_{\rm PS}  &= -E_0\sum_{i=1}^N\Big(\frac{\sigma_i+1}{2} \Big) \\
        &-\frac{K_0}{R}\sum_{i=1}^N \Big(\frac{\sigma_i+1}{2} \Big)\sum_{j=i-R}^{i-1} \Big(\frac{\sigma_j+1}{2} \Big)\\
	&-T\sum_{\mbox{\scriptsize{loops}}}\ln \Big(\Omega \frac{s^l}{l^c} \Big),
    \end{split}
\end{equation}
where $\sigma_i=+1$ and $\sigma_i=-1$ represent bound and open bps, respectively.  Here, the binding energy $E_0$ is assumed to be the same for all bps and 
interaction between bps $K_0$ extends over a range $R$.  In order to have roughly equivalent parameters in both the PDB and PS models, we have set $E_0=D$ and $K_0=K(1+\rho)$.

The final term in Eq.~(\ref{3}) represents the effective potential due to the entropic effects of loops of size $l$.  Using a nearest-neighbor droplet model and the above  
values for $E_0$ and $K_0$, it can be shown that choosing $s=74.4$ leads to a temperature of 350\,K, consistent with our choice in the PDB model.  In nearest-neighbor PS models, the exponent $c$ 
determines the order of the transition. 
We have chosen $c=2.15$, consistent with results for self-avoiding loops published previously~\cite{Blossey}.  The cooperativity $\Omega=0.3$ is larger than values reported elsewhere 
because we wish to consider small loops.

\begin{figure}[h]
\begin{center}
\includegraphics[width=8.cm,height=5.5cm]{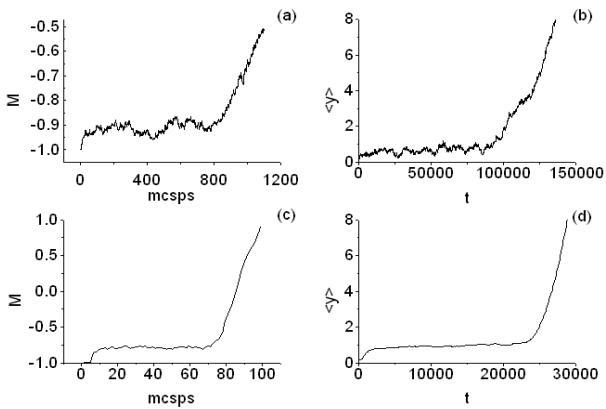}
\caption{\label{fig:ms}Existence of metastable states.  The time evolution is shown for $\lb y\rb$ in the PDB model for (a) PS with $R=1$ and $T=360$, 
(b) PDB with $R=1$ and $T=365$, (c) PS with $R=205$ and $T=517$, and (d) PDB with $R=205$ and $T=685$.}
\end{center}
\end{figure}

We simulate the PDB and PS models with Brownian dynamics (BD) and the Metropolis Monte Carlo (MC) algorithm, respectively.  In BD the system evolves via a Langevin equation 
    \begin{equation}
	\ddot{y_i}=-\nabla V(y_i)-\gamma\dot{y}+\eta(t)
    \end{equation}
where the noise $\eta(t)$ is random Gaussian with $\lb\eta(t)\rb=0$ and $\lb\eta(t)\eta(t^\prime)\rb=2\gamma k_B T \delta(t-t^\prime)$.  
Using a time unit $\tau=1.018\times10^{-14}$\,s, we choose a time step of $0.25\,\tau$ and a damping constant $\gamma=0.1\tau^{-1}$.
In the MC simulations, random spins are flipped and the change in energy between the original and final states is calculated.  Negative changes in energy are always accepted, 
and positive changes are accepted with probability $\exp(-\Delta E/k_B T)$.  For convenience, both simulations use given periodic boundary conditions.

In Figure~\ref{fig:ms} we plot the evolution of both runs for long and short range interactions.  In the PDB model, we monitor the growth of the mean separation of the strands 
$\lb y\rb=\frac{1}{N}\sum y_i$, and for the PS model we observe
the magnetization $M=\frac{1}{N}\sum \sigma_i$.  Each run is instantaneously raised from a low temperature 
to a temperature above the melting temperature.  In PDB we see that for both $R=1$ and $R=205$,
the separation stabilizes after the quench before growing rapidly, indicating that the system enters a metastable state before nucleating.  Likewise, the PS model 
exhibits similar metastability in which the magnetization $M=\frac{1}{N}\sum \sigma_i$ stabilizes before the system nucleates into the stable melted state $M=1$.  

In both the PDB and PS models metastability is observed for all values of $R$ investigated as long as the temperature is not too high.  In general, systems with longer ranges have longer lifetimes and exhibit 
metastability at higher temperatures~\cite{Unger}.  This behavior is due to the fact that the metastable lifetime is proportional to the free energy cost of a critical droplet, which in a one-dimensional system 
scales linearly with the range $R$~\cite{Klein,Unger,Unger2}.  Long-range systems
are closer to mean-field and can be quenched to temperatures close to the mean-field spinodal, while short range systems reach the Becker-D\"{o}ring limit where the system nucleates before 
metastable equilibrium can be achieved.  We have chosen temperatures for each range such that metastability can be observed in a reasonable time.
In both models we find that the nucleation rate decreases with increasing range and that systems with longer ranges can reach metastable equilibrium at higher temperatures than at shorter ranges.   

Nucleation can be viewed as passing from a metastable well to a stable well in a free energy landscape~\cite{Unger}.  
The critical droplet is the system configuration at the top of a saddle point ridge where the 
probability of decaying into the metastable state equals the probability of growing into the stable state.  In order to determine when this droplet occurs during a run, we make 
multiple copies of the system in its initial state and rerun it while adding a random perturbation at some intervention time $t_{\rm int}$~\cite{Gould}.  If after the perturbation the 
system is more likely to decay to the metastable state, then we have not yet reached the critical droplet, while if it tends to nucleate than the droplet is past critical and has entered the growth phase. 
When the probability of returning to 
the metastable state and growing to the stable phase are equal, we say the configuration is the critical droplet.  In the PDB model, we intervene by turning off the noise at 
$t_{\rm int}$ and observing whether the system grows or decays.  In the MC simulations of the PS model, we change the 
random number sequence at $t_{\rm int}$ and rerun the system multiple times to see when the probability of nucleation is roughly one half.  

\begin{figure}[h]
\begin{center}
\includegraphics[width=8.cm,height=5.5cm]{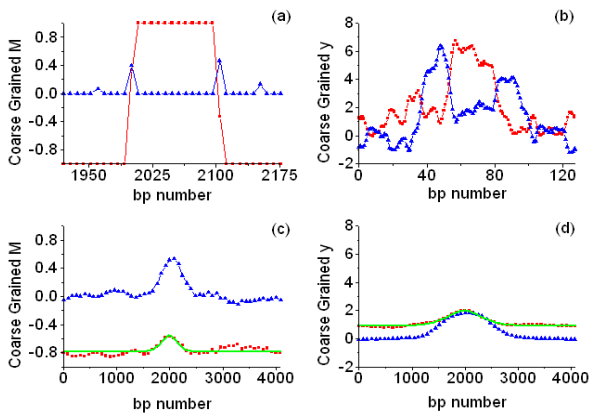}
\caption{\label{fig:CDP} (color online).  Plots of the critical droplet (blue squares), growth mode (red triangles), and the fit (green line) for (a) PS with $R=1$ and $T=360$, 
(b) PDB with $R=1$ and $T=365$, (c) PS with $R=205$ and $T=517$, and (d) PDB with $R=205$ and $T=685$.}
\end{center}
\end{figure}

The profiles of the critical droplets are shown in Figure~\ref{fig:CDP}.   To reduce 
the noise, each profile has been coarse-grained by averaging over all bps within the interaction range.  In both models, smaller $R$ and $T$ give compact 
droplets as expected in classical nucleation.  
Runs with $R>>1$ and high $T$ give diffuse, small amplitude droplet profiles that fit well to an 
inverse hyperbolic cosine squared shape.  These results are consistent with spinodal nucleation~\cite{Unger}.  

We also plot in Figure~\ref{fig:CDP}  the growth modes of the droplet.  Growth modes are obtained by averaging the droplet profiles at 
various times after the critical droplet and 
subtracting the critical droplet profile.  This averaging is done to reduce the noise in the growth mode.  For both models, small $R$ and $T$ gives growth 
modes that are largest at the surface of the droplet, and large $R$ and $T$ systems grow mostly in the center. This difference is expected as classical droplets resemble the
stable phase and grow at the surface, and diffuse spinodal droplets initially grow by filling in at the center~\cite{Unger2}.

In summary, we have observed that nucleation drives the melting transition in modified PDB and PS models of DNA.  We find that long-range interactions, for which there is experimental evidence, 
and deep quench depths give diffuse droplets reminiscent of spinodal nucleation while the short-range shallow quenches give compact classical droplets. These results 
in addition to experimental evidence for long-range interactions \cite{Wartell2} in real DNA suggest the possibility that nucleation in DNA may have spinodal characteristics.

We would like to thank Harvey Gould for useful discussions and both the UNCF-Merck Graduate Fellowship and the DOE for their financial support.

\end{document}